\documentclass[12pt]{iopart}

\usepackage{braket}
\usepackage{graphicx}
\usepackage{subfig}
\usepackage{epstopdf}
\usepackage{iopams}

\newcommand{\ketbra}[3][]{\left|#2\right\rangle_{#1}\!\left\langle#3\right|}

\begin{document}

\title{Maximum-likelihood coherent-state quantum process tomography}
\author{Aamir Anis$^{1}$ and A I Lvovsky$^{2,3}$}

\address{$^1$Indian Institute of Technology, Kharagpur, West Bengal 721302, India}
\address{$^2$Institute for Quantum Information Science, University of Calgary, Calgary AB \\T2N 1N4, Canada}
\address{$^3$Russian Quantum Center, $2^{\rm nd}$ Spasonalivkovsky Pereulok 4, Moscow 121614, Russia}
\ead{lvov@ucalgary.ca}

\begin{abstract}
Coherent-state quantum process tomography (csQPT) is a method of completely characterizing a quantum-optical  ``black box" by probing it with  coherent states and performing homodyne measurements on the output [M. Lobino \etal, Science {\bf 322}, 563 (2008)]. We present a technique for csQPT that is fully based on statistical inference, specifically, quantum expectation-maximization. The method relies on the Jamiolkowski isomorphism and iteratively reconstructs the process tensor in the Fock basis directly from the experimental data. This approach permits incorporation of \emph{a priori} constraints into the reconstruction procedure, thereby guaranteeing that the resulting process tensor is physically consistent. Furthermore, our method is easier to implement and requires a narrower range of coherent states than its predecessors. We test its feasibility using simulations on several experimentally relevant processes.
\end{abstract}

\maketitle

\section{Introduction}

The art of determining states of quantum systems --- quantum tomography --- relies on performing measurements over multiple copies of the state in various bases, followed by reconstruction of the state's density matrix using suitable algorithms on the procured data. Methods of state tomography can be extended to the quantum version of the ``black box" problem \cite{poyatos,dariano,mohseni}, giving rise to quantum process tomography (QPT). In QPT, measurements on the black box response to a certain set of probe states allow one to predict the effect of that black box on any arbitrary state within a given Hilbert space. QPT emerged in response to ever-increasing demands in the field of quantum information processing, as the assembly of any quantum information processor requires precise knowledge of each of its components \cite{fiurasek}.

A popular approach to QPT involves determining the output $\mathcal{E}(\hat{\rho_i})$ for each state of a spanning set $\{\hat{\rho}_i\}$ of the  space of density matrices over the Hilbert space of interest. Due to the linearity of quantum processes over its density operators, the output of any arbitrary state $\hat{\rho}=\sum_i c_i \hat{\rho_i}$ can then be found as $\mathcal{E}(\hat{\rho})=\sum_i c_i \mathcal{E} (\hat{\rho_i})$.

This approach has recently been extended to the continuous-variable domain of quantum optics \cite{lobino}. The reconstruction procedure involves probing the process with coherent states, i.e. simple laser pulses. It relies on the ability of the Glauber-Sudarshan P representation to express the density matrix of  any quantum state as a linear combination of coherent states' density matrices. Improvements in the algorithm have been presented in \cite{saleh}. The algorithm has been tested in an experiment on characterizing quantum-optical memory \cite{memQPT}. Similar principles have recently been used to perform characterization of quantum optical detectors \cite{walmsley1,walmsley2}.

This method, known as coherent-state QPT, or csQPT, has the advantage of employing only the easy-to-prepare coherent states for probing. However, the numerical reconstruction procedures employed in \cite{lobino,saleh}  involve an intermediate step of determining the density matrices of the output states $\mathcal{E}(\ketbra\alpha\alpha)$ for each probe coherent state $\ket\alpha$ and subsequent integration with the P function. This approach requires a multistep calculation and does not guarantee to yield a process that is physically plausible, i.e. completely positive and trace non-increasing.

\begin{figure}
\centering
\includegraphics{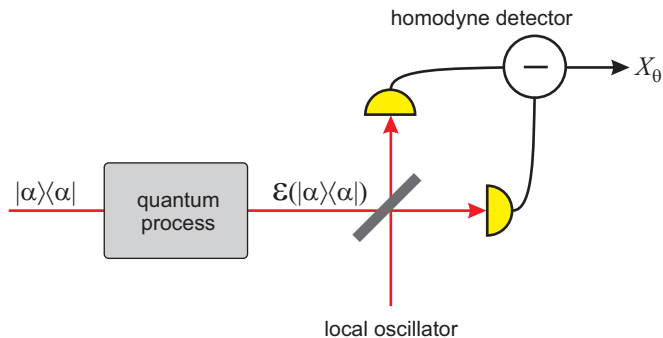}%
\caption{Schematic of the experimental setup for performing csQPT. }
\label{fig:setup}
\end{figure}

We present a reconstruction scheme that does away with this intermediate step, and reconstructs the process directly from the experimental data using pure statistical inference. The experimental setup is equivalent to that of \cite{lobino} and is illustrated in \fref{fig:setup}. The process reconstruction algorithm, on the other hand, is entirely different: it relies on the iterative maximum-likelihood approach. Its major advantage is the possibility to incorporate \emph{a priori} constraints in the reconstruction procedure in order to ensure physically consistent and meaningful results.

Maximum-likelihood methods have been successfully used in the past for quantum state estimation as well as QPT \cite{fiurasek,lvovsky,jezek}. However, their role in QPT has been limited to the discrete variable state space. The technique presented in this paper extends the purview of maximum-likelihood QPT to the continuous variable state space, thereby allowing physically consistent quantum process estimation through homodyne tomography experiments \cite{lvovsky_review}. A further advantage of the present technique is the need of a significantly narrower range of coherent states to probe the process as compared to ~\cite{lobino,saleh}. We test our approach on a number of processes that are relevant to quantum optical information processing: identity, attenuation and photon creation. In doing so, we elaborate a number of recommendations for practical use of the method.

\section{The method}

\subsection{Iterative process estimation using Jamiolkowski isomorphism}

The process reconstruction scheme presented in this paper is based on application of a maximum-likelihood based QPT scheme \cite{fiurasek,jezek} to quadrature measurements in the Hilbert space associated with a harmonic oscillator. Consider a quantum optical process $\mathcal{E}$ acting upon an optical mode prepared in some quantum state $\hat{\rho}_m$. The positivity of density matrices deems it necessary that $\mathcal{E}$ be a completely positive (CP) map, in addition to being trace non-increasing \cite{nielsenchuang}. The output state $\mathcal{E}(\hat{\rho}_m)$ of such a process can be subjected to optical homodyne measurements of its field quadratures $\hat{x}_\theta = \hat{x}\cos\theta + \hat{p}\sin\theta$, where $\hat{x}$ and $\hat{p}$ are the canonical position and momentum operators and $\theta$ is the local oscillator phase. For the output of the probe $\hat \rho_m$, the probability of detecting a specific quadrature value $x$ for a phase $\theta$ is given by
\begin{equation}
p^m_{\theta}(x) = {\rm Tr}\left[ \hat{\Pi}(\theta,x) \mathcal{E}(\hat{\rho}_m) \right],
\label{eq:prob}
\end{equation}
where $\hat{\Pi}(\theta,x) = \ket{\theta,x}\bra{\theta,x}$ is the projector associated with the quadrature eigenstate $\ket{\theta,x}$ and the superscript $m$ on the left hand side denotes the probe state index. The above expression can be considered as a probability distribution function with $\mathcal{E}$ as the parameter. If one performs $N$ measurements for each of the $M$ input probe states $\hat{\rho}_m$, obtained as a set of phase and corresponding quadrature values $\{ \theta_{i,m}, x_{i,m} \}$ where $1\leq i\leq N$ and $1\leq m\leq M$, one can obtain the log-likelihood functional as
\begin{eqnarray}
\mathcal{L}(\mathcal{E}) &= \sum_{m,i} {\rm ln}\left( p^{m}_{\theta_{i,m}}(x_{i,m}) \right) \nonumber \\
&= \sum_{m,i} {\rm ln}\left({\rm Tr}\left[ \hat{\Pi}(\theta_{i,m},x_{i,m}) \mathcal{E}(\hat{\rho}_m) \right]\right).
\label{eq:loglikelihood}
\end{eqnarray}
This functional is convex over the space of CP maps \cite{hradil_book}. The objective of maximum-likelihood estimation is to determine the parameter $\mathcal{E}_{\rm est}$ that is as close to the actual parameter as possible, by maximizing the likelihood functional $\mathcal{L}(\mathcal{E})$ over the space of CP maps
\begin{equation}
\mathcal{E}_{\rm est} = \arg \max_{\mathcal{E}} \mathcal{L}(\mathcal{E}).
\end{equation}
This optimization problem is not straightforward and has been handled previously through various methods  such as the uphill simplex \cite{banaszek}. However, a more rigorous yet technically simpler approach involves the formulation of an extremal equation that maximizes the log-likelihood functional given in equation (\ref{eq:loglikelihood}).

In order to carry out the reconstruction procedure, one needs to first select a certain basis for the representation of the process and the relevant operators. In the Fock (number state) basis, the quantum process can be represented by a rank-4 tensor that relates the density matrix of the input and output states as \cite{lobino,saleh}
\begin{equation}
\left[ \rho_{\rm out} \right]_{jk} = \sum_{m,n,j,k} \mathcal{E}_{jk}^{mn} \left[ \rho_{\rm in} \right]_{mn},
\end{equation}
where $\mathcal{E}_{jk}^{mn} = \braket{j|\mathcal{E}\left(\ket{m}\bra{n}\right)|k}$ and $\rho_{mn} = \braket{m|\hat \rho|n}$. Although the optical Hilbert space is of infinite dimension, in practical process tomography it is truncated to the spanning set of several lowest Fock states, as will be discussed later. Also, the projectors $\hat{\Pi}(\theta,x)$ can be expressed in this basis as
\begin{equation}
\Pi_{mn}(\theta,x) = \braket{m|\hat{\Pi}(\theta,x)|n} = \braket{m|\theta,x}\braket{\theta,x|n},
\end{equation}
where the overlap of the quadrature eigenstate with the number state is given by \cite{lvovsky,leonhardt}
\begin{equation}
\braket{m|\theta,x} = e^{im\theta}\left( \frac{1}{\pi^{1/4}} \right) \frac{H_m(x)}{\sqrt{2^m m!}} e^{-x^2}.
\end{equation}
With the selected basis, we proceed to formulating a numerical procedure for the reconstruction of the quantum process.
For a concise mathematical visualization, we resort to the Jamiolkowski isomorphism between linear CP maps $\mathcal{E}$ from operators on the Hilbert space $\mathcal{H}$ to the space $\mathcal{K}$ and positive semidefinite operators $\hat E$  on the Hilbert space $\mathcal{H}\otimes \mathcal{K}$. The explicit relation between $\hat E$ and $\mathcal{E}$ is given as \cite{hradil_book}
\begin{equation}
\hat E = \sum_{m,n,j,k} \mathcal{E}_{jk}^{mn} \ket{m}\bra{n} \otimes \ket{j}\bra{k}.
\label{eq:Edef}
\end{equation}
With the definition in equation \eref{eq:Edef}, the output $\hat{\rho}_{\rm out} \in \mathcal{K} $ of a process $\mathcal{E}$ for an input $\hat{\rho}_{\rm in} \in \mathcal{H}$ is
\begin{equation}
\hat{\rho}_{\rm out} = \mathcal{E}(\hat{\rho}_{\rm in}) = {\rm Tr_\mathcal{H}} \left[  \hat E  \hat{\rho}_{\rm in}^T \otimes \hat I_\mathcal{K} \right],
\label{eq:rho_out}
\end{equation}
where $T$ denotes transposition in the number basis. In addition, one must apply the trace-preservation condition (${\rm Tr}[\hat{\rho}_{\rm out}] = {\rm Tr}[\hat{\rho}_{\rm in}]$) over the process $\mathcal{E}$, which yields
\begin{equation}
{\rm Tr_\mathcal{K}} [\hat E] = \hat I_\mathcal{H}.
\label{eq:trace}
\end{equation}
The reconstruction procedure can be extended to also encompass trace non-preserving processes, as will be shown subsequently. The problem has thus reduced to the determination of $({\rm dim}\mathcal{H}{\rm dim}\mathcal{K})^2$ parameters subject to ${\rm dim}\mathcal{H}^2$ constraints. When the input and output Hilbert spaces are identical, this amounts to evaluating ${\rm dim}\mathcal{H}^4-{\rm dim}\mathcal{H}^2$ free parameters.

For the process output of the input probe state $\hat{\rho}_m$, the probability of reading a quadrature value $x$ for a given local oscillator phase $\theta$ can be obtained by substituting \eref{eq:rho_out} into equation~\eref{eq:prob} to obtain
\begin{equation}
p^m_{\theta}(x) = {\rm Tr}\left[  \hat E \hat{\rho}_m^T \otimes \hat{\Pi}(\theta,x) \right].
\label{eq:pdf}
\end{equation}
Operator $\hat E$ should then maximize a constrained log-likelihood functional in order to stand as the most likely quantum process that has the set of outcomes  $\{ \theta_{i,m}, x_{i,m} \}$ for the input probes $\{ \hat{\rho}_m \}$. The relevant log-likelihood functional is given as
\begin{eqnarray}
\mathcal{L}(\hat E) &= \sum_{m,i} {\rm ln}\left( p^{m}_{\theta_{i,m}}(x_{i,m}) \right) - {\rm Tr}[\hat \Lambda \hat E] \nonumber \\
&= \sum_{m,i}{\rm ln}\left( {\rm Tr}\left[ \hat E \hat{\rho}_m^T \otimes \hat{\Pi}(\theta_{i,m},x_{i,m}) \right] \right) - {\rm Tr}[\hat \Lambda \hat E],
\label{eq:likelihoodeqn}
\end{eqnarray}
where $\hat \Lambda = \hat \lambda \otimes \hat I_\mathcal{K}$ and $\hat \lambda$ is the Hermitian matrix of Lagrange multipliers that incorporates the trace preservation condition (\ref{eq:trace}). Again, $\theta_{i,m}$ and $x_{i,m}$ belong to the set of quadrature data for the $m^{{\rm th}}$ probe state given by $\{\theta_{i,m},x_{i,m}\}$. An extremal equation can be obtained by varying equation~\eref{eq:likelihoodeqn} with respect to $\hat E$:
\begin{equation}
\delta \mathcal{L}(\hat E) = \mathcal{L}(\hat E+\delta \hat E) - \mathcal{L}(\hat E) = 0,
\end{equation}
which gives
\begin{equation}
{\rm Tr}\left[ \left( \sum_{m,i} \frac{1}{p^m_{\theta_{i,m}}(x_{i,m})} \hat{\rho}_m^T \otimes \hat{\Pi}(\theta_{i,m},x_{i,m}) - \hat \Lambda \right) \delta \hat E \right] = 0.
\end{equation}
This holds for all  $\delta \hat E$, so that the expression in the parentheses can be equated to zero  and one has
\begin{equation}
\label{eq:E}
\hat E = \hat \Lambda^{-1} \hat R \hat E,
\end{equation}
where
\begin{equation}
\label{eq:R}
\hat R = \sum_{m,i} \frac{1}{p^m_{\theta_{i,m}}(x_{i,m})} \hat{\rho}_m^T \otimes \hat{\Pi}(\theta_{i,m},x_{i,m}).
\end{equation}
Owing to Hermicity, one may rewrite equation (\ref{eq:E}) as $\hat E = \hat E \hat R \hat  \Lambda^{-1}$. Using this, along with equation (\ref{eq:E}),  we arrive at
\begin{equation}
\hat E = \hat \Lambda^{-1}\hat R \hat E \hat R\hat \Lambda^{-1}.
\label{eq:finalE}
\end{equation}
$\hat \Lambda$ can be determined by substituting the expression for $\hat E$ in equation~\eref{eq:finalE} into the trace-preservation condition (\ref{eq:trace}):
\begin{equation}
\hat \lambda = \left( {\rm Tr_\mathcal{K}}[\hat R \hat E \hat R] \right)^{1/2}.
\label{eq:lambda}
\end{equation}
Equations (\ref{eq:finalE}) and (\ref{eq:lambda}) can be solved numerically through iterations, starting from an unbiased initial $\hat E$, such as $\hat E^{(0)} = \hat I_{\mathcal{H}\otimes \mathcal{K}}/({\rm dim\mathcal{K}})$. At each step of the iterations, the positive semi-definiteness of $\hat E$ is ensured and the constraint ${\rm Tr_\mathcal{K}}[\hat E] = \hat I_\mathcal{H}$ is satisfied.

Quantum processes may also be probabilistic, in which case the trace of the input quantum state is not preserved. The probability of occurrence of a probabilistic quantum process is given by
\begin{equation}
p_{\rm success} = {\rm Tr}[\mathcal{E}(\hat{\rho})],
\end{equation}
The reconstruction of probabilistic quantum processes can be viewed as a reconstruction of a trace-preserving, deterministic CP map $\tilde{\mathcal{E}}$ if the failure of the process is taken to be a measurement event associated with the projection operator $\hat \Pi_\emptyset$ onto a fictitious state  $\ket{\emptyset}$ \cite{hradil_book}. In order to analyze such a process, one can extend the Hilbert space to form $\mathcal{K}_{\rm total} = \mathcal{K} \oplus \mathcal{K}_{\rm fail}$, where $\mathcal{K}_{\rm fail}$ is spanned by the single state  $\ket{\emptyset}$. The original set of projectors $\hat{\Pi}_\theta(x)$ for each $\theta$ is augmented by adding $\hat{\Pi}_\emptyset$ so that the new set of projectors satisfies the closure relation over $\mathcal{K}_{\rm total}$, i.e. $\forall \theta\int \hat{\Pi}_\theta(x) dx + \hat{\Pi}_\emptyset = I$. Subsequently, the likelihood functional, with the extended trace-preserving map $\tilde{\mathcal{E}}$ as parameter, can be rewritten as
\begin{equation}
\label{eq:heralded}
\mathcal{L}(\hat{\tilde{E}}) =  \sum_{m,i}\left[g_m {\rm ln}\left( p^{m}_{\theta_{i,m}}(x_{i,m}) \right) + (1-g_m) {\rm ln}\left( p^{m}_\emptyset \right) \right] - {\rm Tr}[\hat \Lambda \hat{\tilde{E}}],
\end{equation}
where $g_m$ is the fraction of successful events over total events, which can be determined experimentally. The extremal equation would then contain a modified operator $\hat R$ given by
\begin{equation}
\hat{\tilde{R}} = \sum_{m,i} \left[ \frac{g_m}{p^m_{\theta_{i,m}}(x_{i,m})} \hat{\rho}_m^T \otimes \hat{\Pi}(\theta_{i,m},x_{i,m})  +   \frac{1-g_m}{p^m_\emptyset} \hat{\rho}_m^T \otimes \hat{\Pi}_\emptyset \right].
\end{equation}
Iterations can now be performed with the new $\hat{\tilde{R}}$ to obtain the trace-preserving process tensor $\hat{\tilde{E}}$. The actual process tensor $\hat E$ is obtained by taking the projection of the estimated tensor $\hat{\tilde{E}}$ onto the subspace $\mathcal{H}\otimes \mathcal{K}$.

Our analysis so far did not specify which states were to be used as probes; the only requirement is that these states compose a spanning set in the space of density matrices. In csQPT, the role of probe states is played by coherent states \cite{lobino}. The density operator of an arbitrary state can be written as a linear combination of coherent state density operators using the optical equivalence theorem:
\begin{equation}
\hat\rho = \int P_{\hat\rho}(\alpha) \ket{\alpha}\bra{\alpha} d^2\alpha,
\end{equation}
where $P_{\hat\rho}(\alpha)$ is the Glauber-Sudarshan P function of state $\hat\rho$.
Using the linearity of quantum processes with respect to density matrices, the process output is then given by
\begin{equation}
\mathcal{E}(\hat\rho) = \int P_{\hat\rho} \mathcal{E}(\ket{\alpha}\bra{\alpha}) d^2\alpha.
\end{equation}
Therefore, if the response of the quantum system to all coherent states is known, the output of any arbitrary unknown quantum state can be computed. In other words, measurements on the set of responses $\mathcal{E}(\ket{\alpha_m}\bra{\alpha_m})$ for coherent states $\ket{\alpha_m}$ provides tomographically complete information about the quantum process.

\subsection{Practical issues}
\label{pracSec}

We now proceed to discussing a few practical issues arising in the implementation of the above algorithm of csQPT.
The first issue is associated with infinite dimension of the optical Hilbert space. In practical implementation of csQPT, the process tensor is reconstructed for a subspace $\mathbb{H}(n_{\max})$ of the Hilbert space spanned by Fock states up to a certain cut-off value, $n_{\rm max}$. The choice of $n_{\rm max}$ is correlated with the maximum amplitude  $\alpha_{\rm max}$ of the set of coherent probe states. Given a data set with a specific $\alpha_{\rm max}$, the choice of $n_{\rm max}$ depends on many factors, in particular, the process itself (see supplementary online material to \cite{lobino}).

For the iterative cycle, the cut-off value must be chosen sufficiently high so that $\mathbb{H}(n_{\max})$  accommodates all of the coherent probe states and the associated output states. Otherwise, the probe states and the quadrature data will be inadequately represented by $\mathbb{H}(n_{\max})$. This will lead to inaccurate reconstruction of the process tensor; we refer to this phenomenon as \emph{truncation errors}.

For physically realistic processes, we expect the fractions of $\ket{\alpha_m}$ and $\mathcal{E}(\ket{\alpha_m}\bra{\alpha_m})$ that lie outside the reconstruction subspace to vanish as $n_{\max}$ tends to infinity. Hence, for a given $\alpha_{\max}$, it is possible to choose a value of $n_{\max}$ such that the associated truncation errors are arbitrarily low \cite{lutk11}.

However, a high cut-off value may give rise to another class of inaccuracies, which we call \emph{data insufficiency errors}. If the overlap of a given Fock state $\ket n$ with all of the $\ket{\alpha_m}$ is low, so is the contribution of $\ket n$ to the log-likelihood functional, and hence the available data will not provide sufficient information about the effect of the process on $\ket n$. In contrast to the truncation errors, the data insufficiency errors grow with $n_{\rm max}$, but only apply to the process tensor elements associated with high input photon numbers.

Therefore the following dual-step procedure for the choice of the cut-off may be necessary. The initial value  of  $n_{\rm max}$ must be sufficiently high to ensure absence of truncation errors. Subsequently, after the iterative cycle has been completed, we choose a secondary cut-off value, $n'_{\rm max}\le n_{\rm max}$, and remove all the process tensor elements containing indices above $n'_{\max}$. The choice of $n'_{\max}$ can be determined by calculating the statistical errors associated with each process tensor element - similar to the error estimations for state tomography \cite{hradil_book,rehacek_error,blume-kohout,renner}. However, further research is required to determine statistical errors for QPT and establish a concrete bound for $n'_{\rm max}$. In the next section, we illustrate the effect of the chosen subspace dimension on the process reconstruction through various simulations.

As in the case of state tomography \cite{lvovsky}, our algorithm permits automatic correction for optical losses and inefficient detectors in the  process tensor reconstruction. In order to account for non-unitary efficiency $\eta$, the projection operators are replaced by
\begin{equation}
\fl \hat{\Pi}_\eta(\theta,x) = \sum_{m,n,j,k} B_{n+k,n}(\eta)B_{m+k,m}(\eta) \braket{m|\hat{\Pi}(\theta,x)|n} \times \ket{m+k}\bra{n+k},
\end{equation}
where $B_{n+k,n} = \left[{n+k \choose k}\eta^n(1-\eta)^k\right]^{1/2}$. Substituting this into equation~\eref{eq:R} and performing the iterations generates the original process tensor pertaining to the case of ideal detection.

Many physically relevant processes are phase-invariant: applying an optical phase shift to the input state results in the same shift to the output. Mathematically, such processes satisfy the following relation \cite{saleh,lvovsky}
\begin{equation}
\mathcal{E}[\hat{U}(\phi)\hat{\rho}\hat{U}^\dagger(\phi)] = \hat{U}(\phi)\mathcal{E}(\hat{\rho})\hat{U}^\dagger(\phi)
\label{eq:phase_invariance}
\end{equation}
In this case,  further simplifications can be made. If the action of the process on a coherent state $\ket{\alpha}$ is known, so is the outcome for $\ket{\alpha e^{i \phi}}$. Therefore, one needs to only perform measurements for input coherent states with amplitudes on the positive real axis. When condition (\ref{eq:phase_invariance}) is applied in the Fock basis, the elements of the process tensor $\mathcal{E}^{mn}_{jk}$ for a phase-invariant process vanish except when $m-n=j-k$. This condition is incorporated into the probability distribution (\ref{eq:pdf}) as
\begin{equation}
\label{eq:newpdf}
p^m_{\theta}(x) = {\rm Tr}\left[  \mathcal{M}(\hat E) \hat{\rho}_m^T \otimes \hat{\Pi}(\theta,x) \right],
\end{equation}
where $\mathcal{M}$ denotes a masking operation over $\hat E$. If $\hat \Pi_m = \ket{m}\bra{m}$ denotes a projection operator in the number basis, then $\mathcal{M}(\hat E)$ can be expressed as
\begin{equation}
\mathcal{M}(\hat E) = \sum_{m,n,j,k}\delta_{m-n,j-k}(\hat \Pi_m\otimes \hat \Pi_j) \hat E (\hat \Pi_n \otimes \hat \Pi_k).
\end{equation}
Since the trace operation is invariant under cyclic rearrangements of the operators and the Kronecker delta is invariant under transposition of indices, the probability distribution (\ref{eq:newpdf}), and consequently, the expression for the operator $\hat R$ in equation \eref{eq:R} changes to
\begin{eqnarray}
p^m_{\theta}(x) = {\rm Tr}\left[ \hat E \mathcal{M}\left(\hat{\rho}_m^T \otimes \hat{\Pi}(\theta,x)\right) \right] \\
\hat R = \sum_{m,i} \frac{1}{p^m_{\theta_{i,m}}(x_{i,m})} \mathcal{M}\left( \hat{\rho}_m^T \otimes \hat{\Pi}(\theta_{i,m},x_{i,m}) \right).
\end{eqnarray}
When the above relations are used, the elements of $\mathcal{E}_{jk}^{mn}$,  for which $m-n\neq j-k$, vanish, resulting in the incorporation of the phase invariance condition.

In some cases, the value of the log-likelihood oscillates before converging to the maximum owing to overshoots. Stabilization can be achieved using the diluted algorithm that slows down but guarantees convergence \cite{rehacek}. The operator $\hat R$, in that case, is modified to a weighted sum of itself and the identity operator as
\begin{equation}
\hat R' = \mu \hat R +  (1-\mu) \hat I,
\label{eq:diluted}
\end{equation}
where $0\leq \mu \leq 1$. As the value of $\mu$ decreases, the algorithm becomes more and more dilute, resulting in increased stability but a reduced rate of convergence. In addition, monotonic increase of the likelihood is guaranteed for small values of $\mu$ (see \ref{sec:proof_convergence}). One may try to find the optimal value of $\mu$ that maximizes the increase in likelihood at the cost of increased computational complexity. Gradually varying the value of $\mu$ during the iterations may be justified for some processes.

The number of quadrature measurements for each probe state typically ranges in tens of thousands. With multiple probe states, the iteration cycle may require significant computation time. In order to speed up the computation, binning of the data points in the quadrature and phase axes may be useful. A suitable step size is chosen for each axis as a trade-off between the desired computational time and the quantization error. For each $\mathcal{E}(\ket{\alpha_m}\bra{\alpha_m})$, quadrature data points are then clubbed into bins with centers $\{\theta_{u,m},x_{v,m}\}$. With this modification, the log-likelihood functional~\eref{eq:loglikelihood} now reads as
\begin{equation}
\label{eq:freq1}
\mathcal{L}(\mathcal{E}) = \sum_{m,u,v}h_{m;u,v}{\rm ln}\left[p^{m}_{\theta_{v,m}}(x_{u,m}) \right],
\end{equation}
where $h_{m;u,v}$ denotes the number of data points in the bin with center $(\theta_{u,m},x_{v,m})$. Ideally, one must obtain the POVM associated with the bin center as a function of all the POVMs lying in the bin. However, given a small size of the bin, this element can be approximated by projection onto the quadrature value at the center of the bin. Similarly, the operator $\hat R$ in equation~\eref{eq:R} can be rewritten as
\begin{equation}
\label{eq:freq2}
\hat R = \sum_{m,u,v} \frac{h_{m;u,v}}{p^m_{\theta_{v,m}}(x_{u,m})} \hat{\rho}_m^T \otimes \hat{\Pi}(\theta_{v,m},x_{u,m}).
\end{equation}
For further speedup, one can compute $\hat R$ in a parallel fashion on different threads owing to absence of interdependency in the summation procedure.

In practical experiments on probabilistic processes, the frequency of successful events can be low: $g_m \ll 1$. In this case, the process tensor elements of interest (i.e. those related to $\mathcal{K}$) will be small and thus suffer from increased relative error. This issue can be resolved by rescaling the values of all $g_m$ by the same factor for all probe states, keeping in mind the requirement that $g_m<1$ for all $m$. Physically relevant elements of the process tensor will then rescale by the same factor, reducing the relative error.

\section{Implementation and results}

\begin{figure*}
\centering
  \subfloat[]{\includegraphics[width=0.3\linewidth]{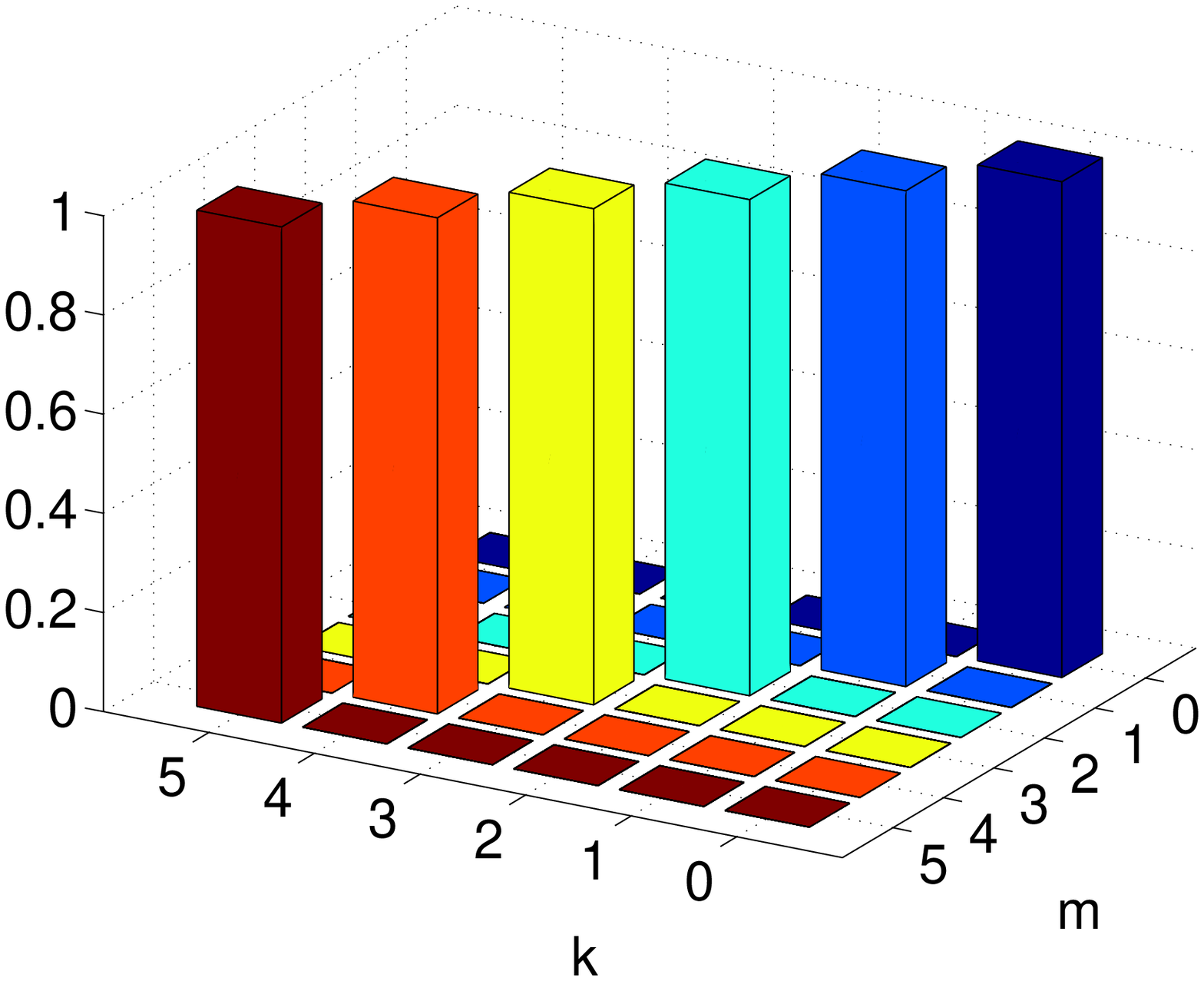}} \quad
  \subfloat[]{\includegraphics[width=0.3\linewidth]{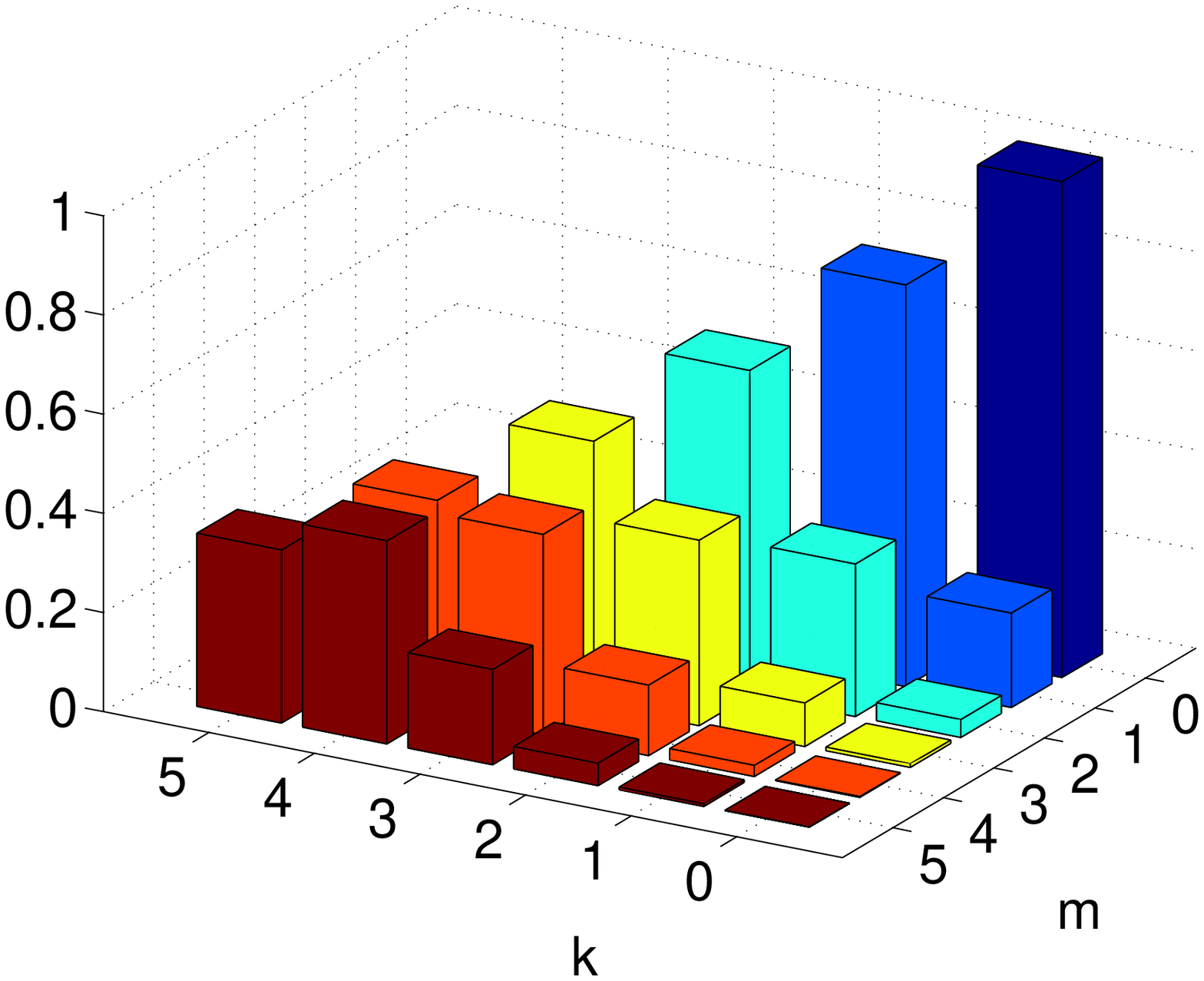}} \quad
  \subfloat[]{\includegraphics[width=0.3\linewidth]{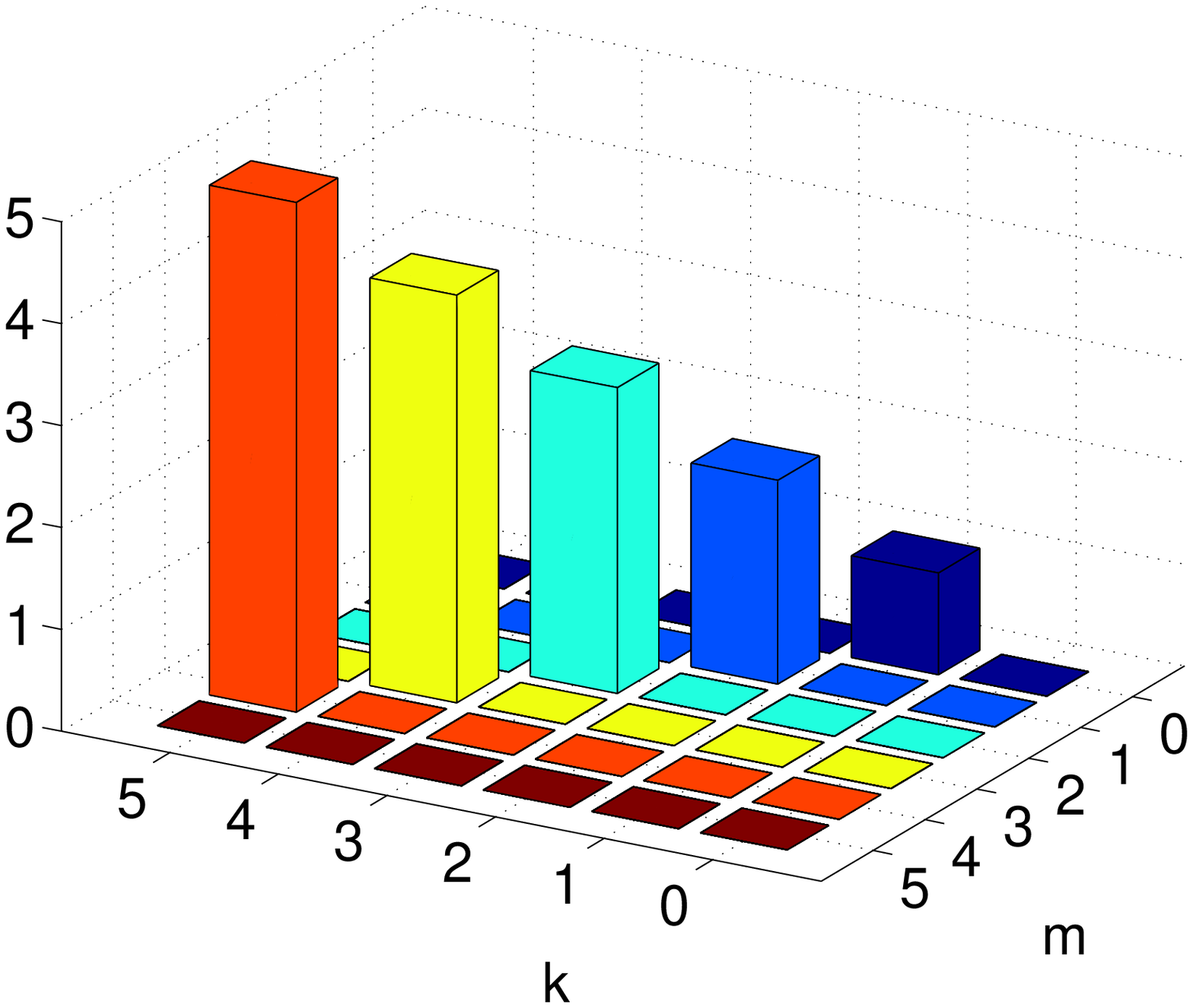}} \\
  \subfloat[]{\includegraphics[width=0.3\linewidth]{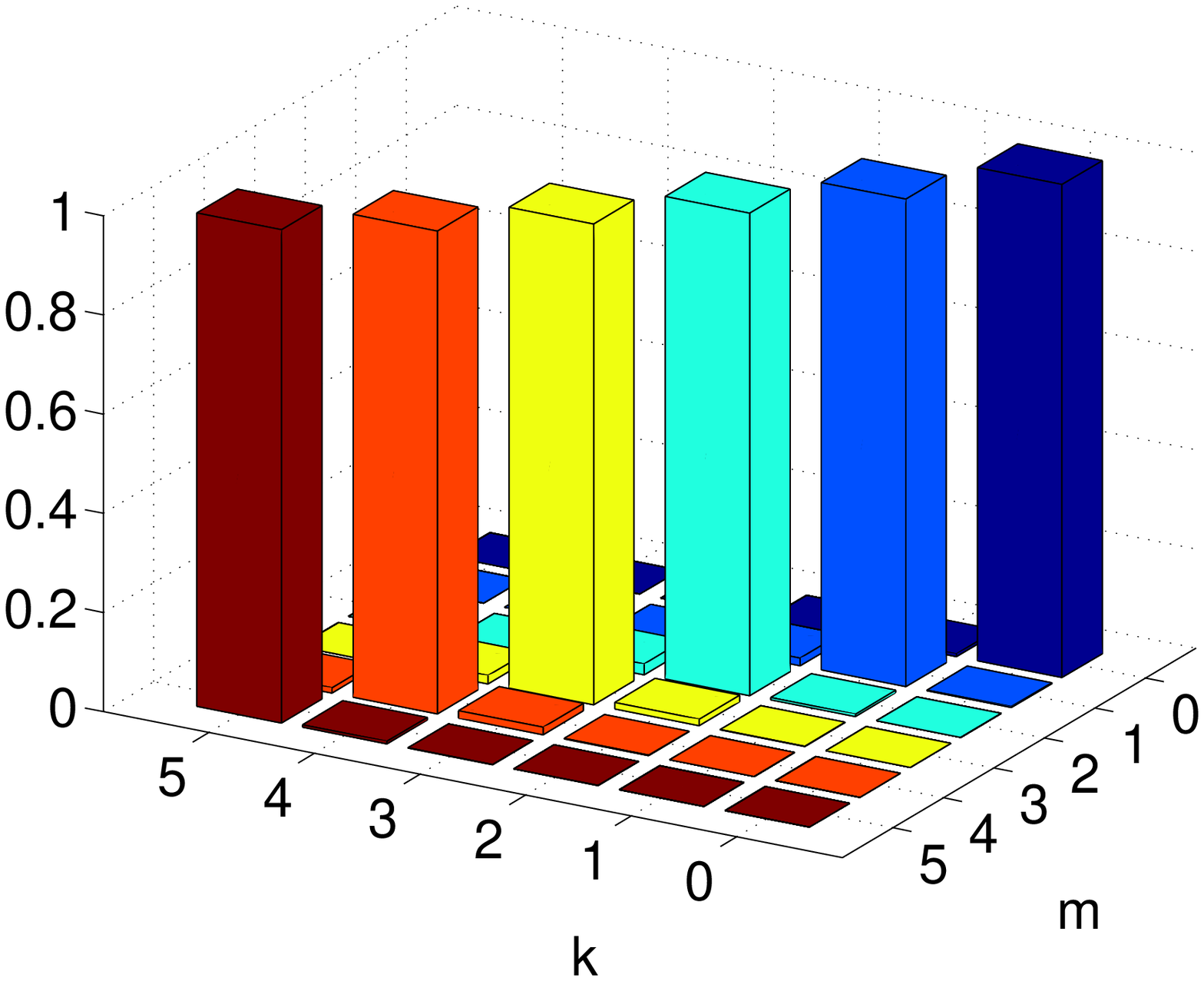}} \quad
  \subfloat[]{\includegraphics[width=0.3\linewidth]{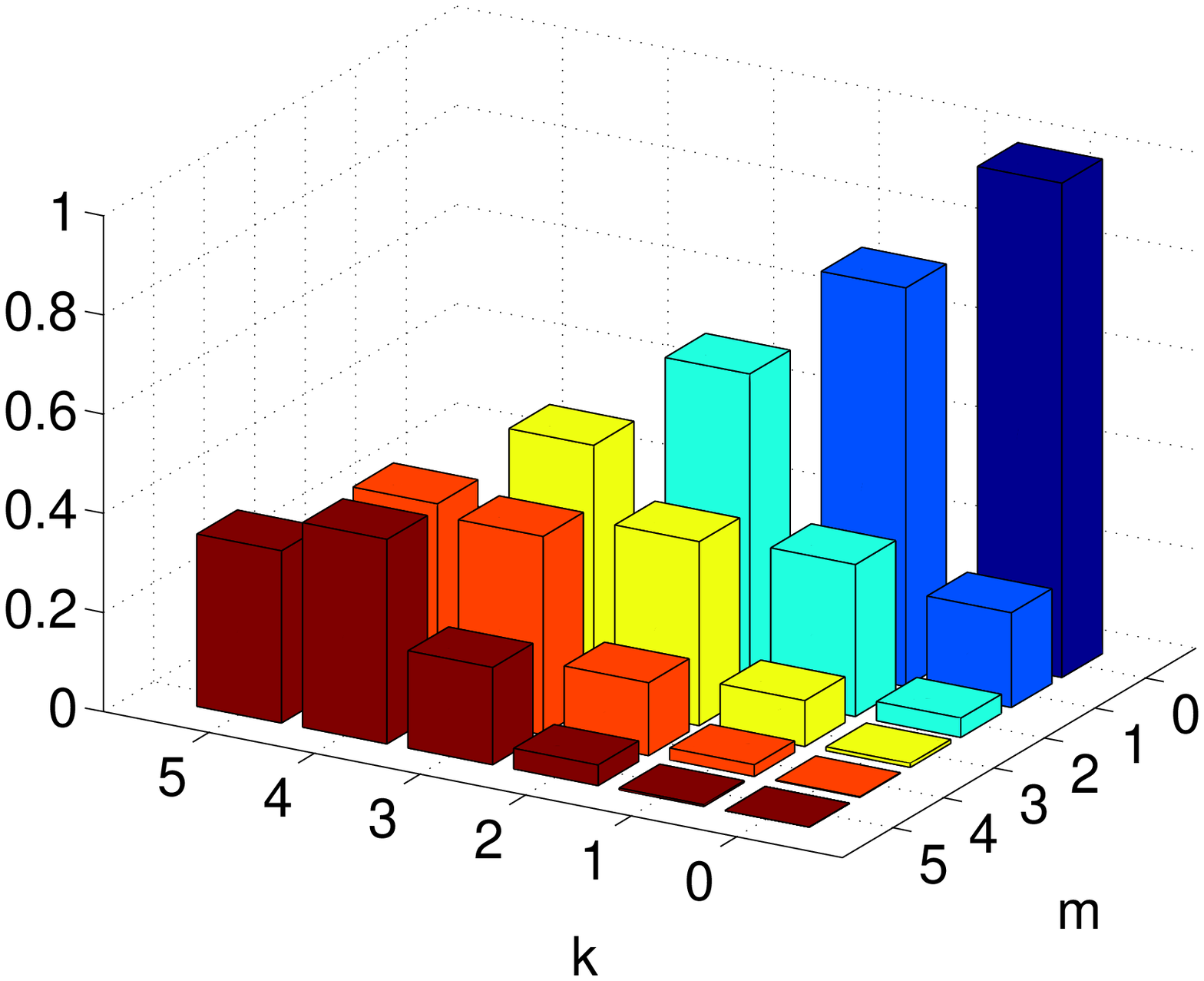}} \quad
  \subfloat[]{\includegraphics[width=0.3\linewidth]{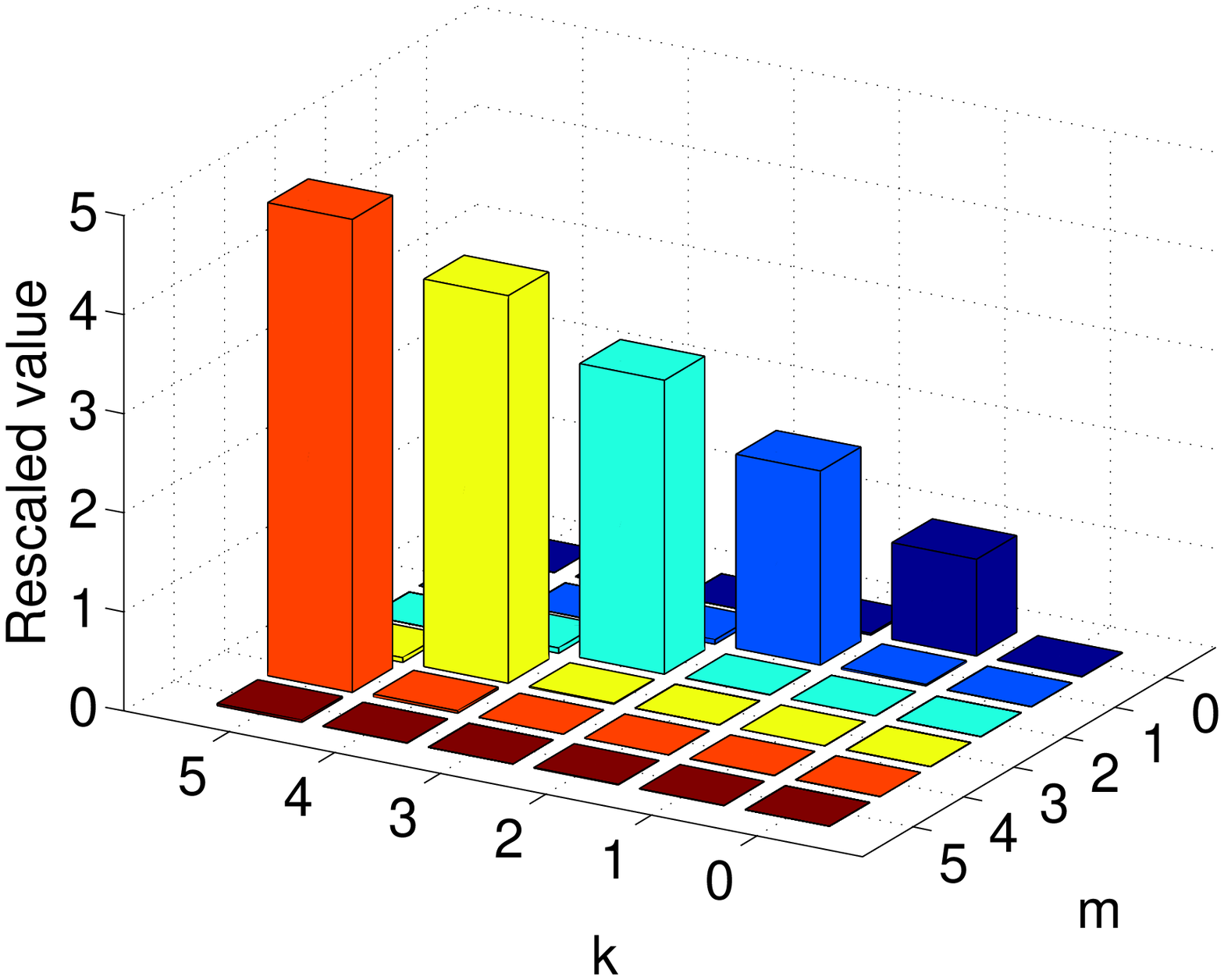}}
  \caption{Comparison of the diagonal values ($\mathcal{E}^{mm}_{kk}$) of theoretical and reconstructed process tensors. (a), (b) and (c) show process tensors for theoretical identity, attenuation (by factor $0.9$) and photon creation processes, respectively. (d), (e) and (f) show the process tensors for the corresponding reconstructed processes using the algorithm presented in this article. The photon creation process tensor has been scaled to match the theoretical one (i.e. $g=1$).}
  \label{fig:results}
\end{figure*}

In order to test the algorithm, we have implemented it using Matlab and studied the reconstruction of a few quantum processes using simulated data. Theoretical process tensors of identity, attenuation and photon creation  \cite{saleh} were used to find the marginal probability distribution functions for various probe states using equation~\eref{eq:pdf}. From the marginals, we generated synthetic experimental data through Monte-Carlo simulations.

Each process was applied to four coherent probe states with $\alpha_m$ ranging from 0 to 0.9375 in steps of 0.3125. For each input probe state, the output state dataset consisted of 100,000 phase and quadrature points $\{\theta,x\}$. This set of data was subjected to the reconstruction method described. The iterations were halted when the change in process tensor elements was insignificant over a large number of iterations. However, a better approach would be to set a threshold for the increase of the log-likelihood \cite{knill}.

The result obtained by running the reconstruction technique is a 4-dimensional process tensor whose diagonal elements have a simple interpretation. For a given quantum process $\mathcal{E}$, the diagonal element $\mathcal{E}^{mm}_{kk}$ denotes the probability that the output contains $k$ photons when the process is subjected to $m$ input photons. A comparison between the diagonal elements of the theoretical and reconstructed process tensor is made in \fref{fig:results} and exhibits close match between the two.

The process of photon creation $\hat{a}^\dagger$ requires additional discussion because it corresponds to a non-unitary, trace non-preserving operator. Therefore, in experimental practice it can only be implemented probabilistically. The optical mode containing the target state $\ket{\psi}$ is directed into the signal channel of a parametric down-conversion setup. The state of the down-conversion output in the signal ($s$) and idler ($i$) channels can then be written as $\ket{\psi}_s\ket{0}_i + g(\hat{a}^\dagger \ket{\psi})_s\ket{1}_i$, where $g$ is the down-conversion amplitude. Detection of a photon in the idler channel projects the signal state onto $\hat{a}^\dagger \ket{\psi}$, thereby heralding a photon addition event \cite{zavatta}. For coherent state input $\ket\psi=\ket\alpha$, the event probability, corresponding to the quantity $g_{m}$ in equation~\eref{eq:heralded}, is proportional to $|g|^2(1+|\alpha|^2)$.

We take the value $g^2=0.1$ during simulations to ensure that success probabilities remain less than 1 for the probe states selected. This makes the photon creation process trace non-increasing and thus physical. Note that the process tensor reported in \fref{fig:results}(f) has been normalized by dividing by $g^2$, so that its scale matches that of the process tensor for the photon creation operator $\hat a^\dag$ given in \fref{fig:results}(c).

\begin{figure*}
  \centering
  \subfloat[]{\includegraphics[width=0.3\linewidth]{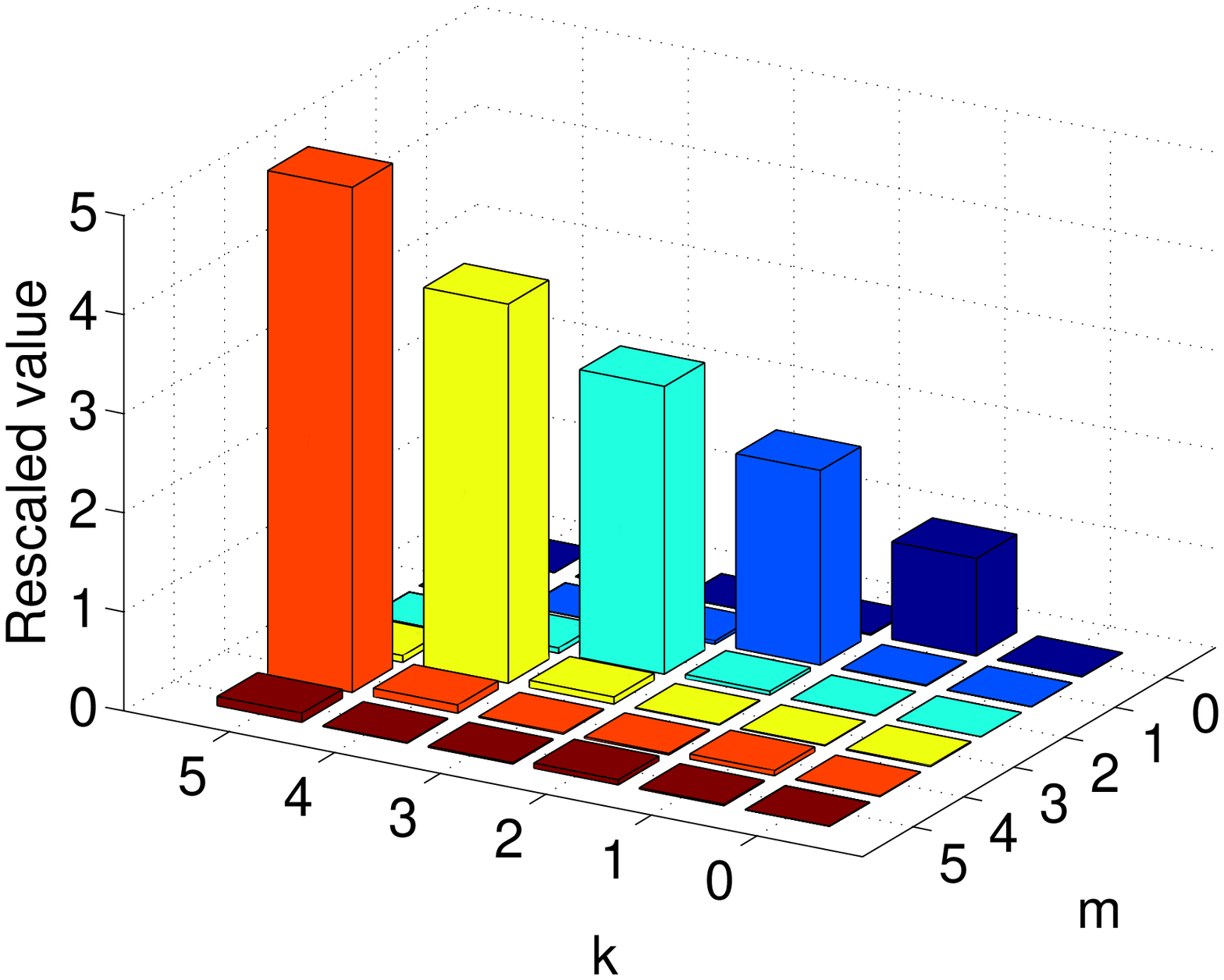}} \quad
  \subfloat[]{\includegraphics[width=0.3\linewidth]{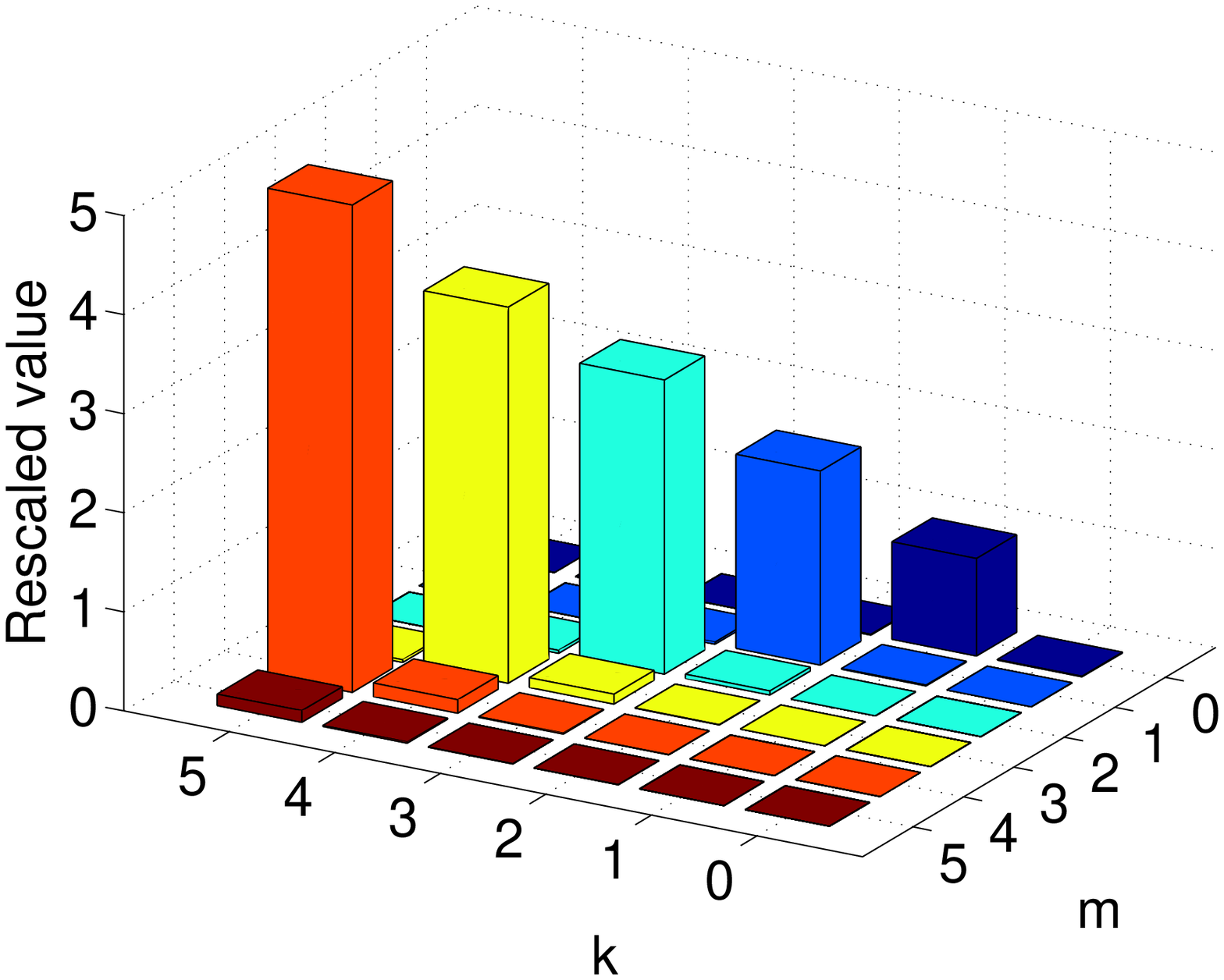}}
  \caption{Reconstruction of the photon creation process with correction for inefficiency: (a) $\eta=0.75$ and (b) $\eta=0.55$.}
  \label{fig:inefficient}
\end{figure*}

The iterative reconstruction of photon creation exhibited relatively poor convergence. Diluted iterations (\ref{eq:diluted}) were required in the beginning in order to curb oscillations. However, as the iterations progressed, the rate of increase of the likelihood value became extremely low. We circumvented this issue by implementing the successive over-relaxation technique. Setting $\mu$  in equation \eref{eq:diluted} to slightly over $1$ while iterating  accelerated the increase in likelihood. As soon as a decrease in the likelihood value was registered due to an overshoot, $\mu$ was reset to $1$, and then slowly increased again after stabilization. This procedure was applied multiple times until a fair amount of convergence was observed. In a loose sense, the over-relaxation method employs linear extrapolation by selecting a tensor that lies on the line joining the current iterate and the next iterate but is beyond the latter by a fraction. If the iterations happen to proceed in the direction of maximum likelihood gradient, it allows faster convergence by inducing greater leaps. Additionally, it may also help in escaping limit cycles encountered during the iterations.

We have also tested the reconstruction technique for photon creation in the case of inefficient detection. The output density matrices for the probe states were calculated using the beam splitter model of absorption \cite{leonhardt}.  With these modified density matrices, we have generated test data using Monte Carlo simulations for $\eta=0.75$ and $\eta=0.55$. A comparison of the reconstructed process tensors is given in \fref{fig:inefficient}.

Finally, we investigated the effect of the dimension of the subspace of optical Hilbert space chosen for the reconstruction, specifically for the photon creation process. In order to eliminate statistical errors in this reconstruction, we directly used the marginal distributions  instead of simulated quadrature data sets to obtain the values of $h_{m;u,v}$ in equations \eref{eq:freq1} and \eref{eq:freq2}. The performance criterion is taken to be the worst-case fidelity \cite{lvovsky_memory_review} over the input space $\mathbb{H}(n_{\max}-1)$, defined as
\begin{equation}
\mathcal{F}(\mathcal{E},\mathcal{E}_{\rm est}) = \min_{\hat{\rho} \in \mathbb{H}(n_{\max}-1)} {\rm Tr}\left( \sqrt{ \sqrt{\mathcal{E}(\hat{\rho})} \; \mathcal{E}_{\rm est}(\hat{\rho}) \; \sqrt{\mathcal{E}(\hat{\rho})} } \right),
\end{equation}
where $\mathcal{E}$ is the actual process tensor and $\mathcal{E}_{\rm est}$ is the estimated process tensor. Note that the photon number cutoff for the fidelity calculation is taken to be $n_{max}-1$ to ensure the Hilbert space closure under photon addition. Minimization over $\mathbb{H}(n_{\max}-1)$ is carried out through a Monte Carlo simulation that involves introducing small random changes in the density matrix $\hat{\rho}$ within $\mathbb{H}(n_{\max}-1)$ and accepting the change whenever the value of the fidelity decreases.

\begin{figure*}
\centering
 \subfloat[]{\includegraphics[width=0.36\linewidth]{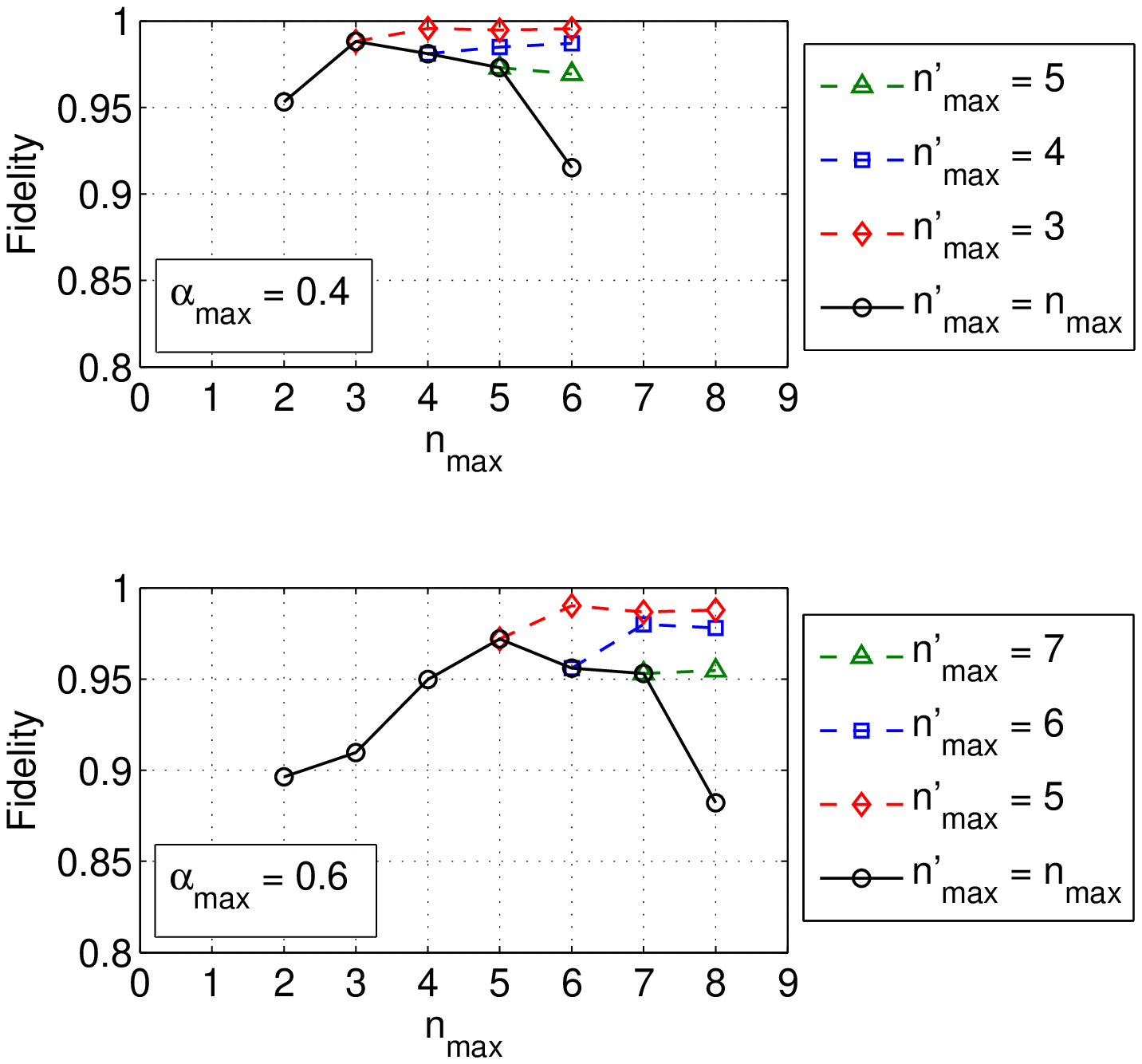}} \quad
  \subfloat[]{\includegraphics[width=0.20\linewidth]{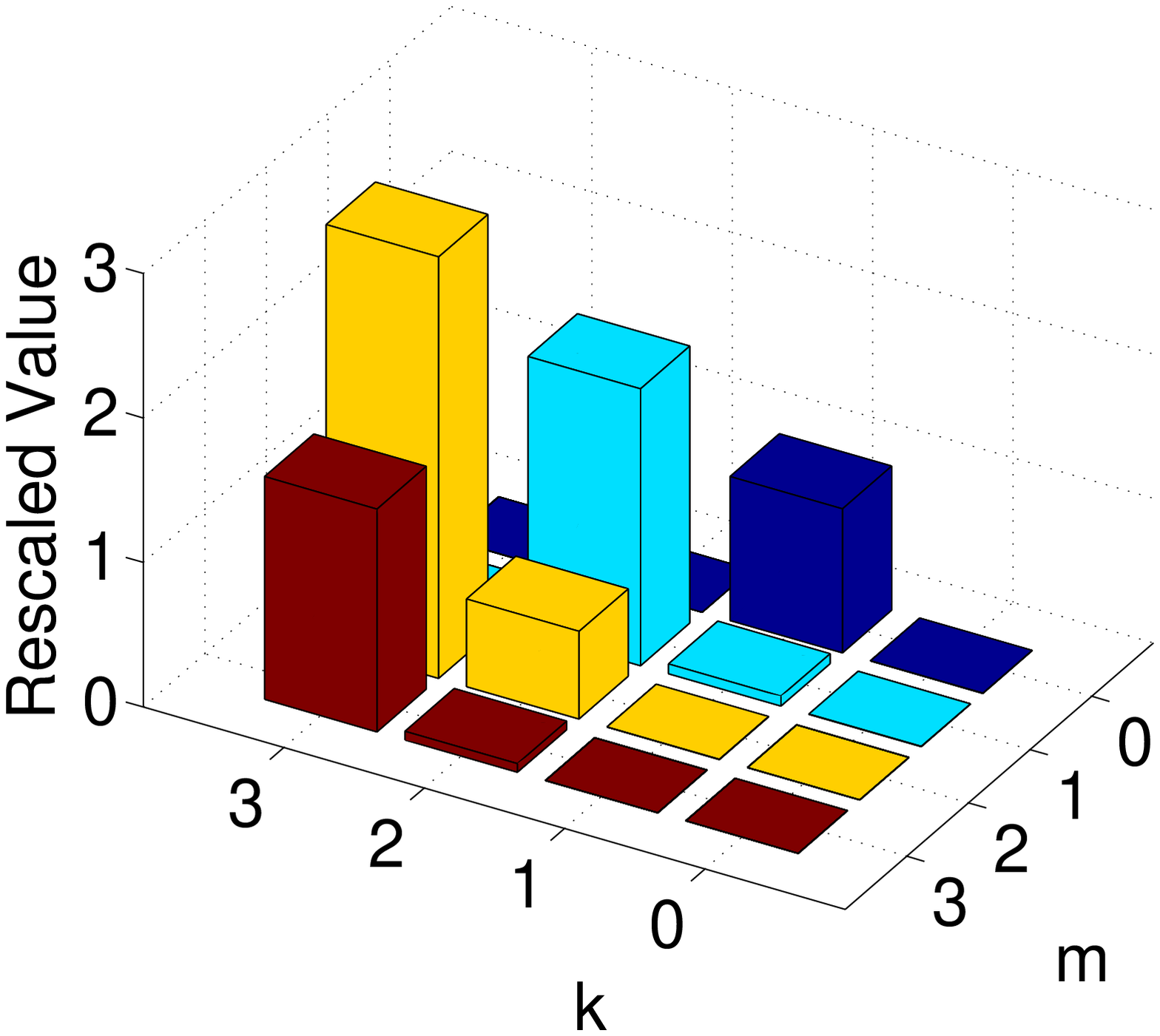}} \quad
  \subfloat[]{\includegraphics[width=0.36\linewidth]{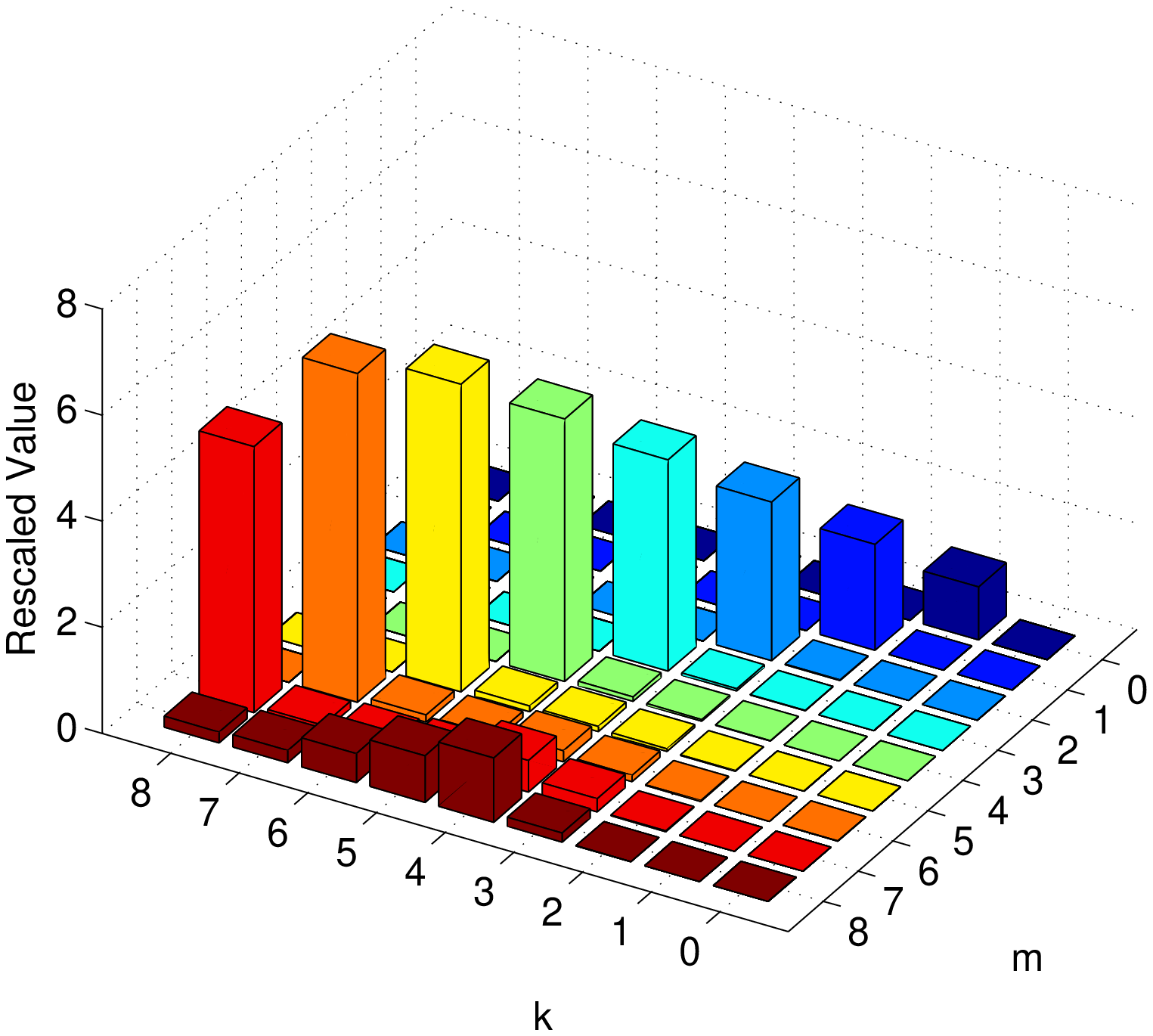}}
  \caption{Effect of the photon number cut-off ($n_{\rm max}$) on the reconstruction of the photon creation process. (a) Worst-case process reconstruction fidelity as a function of $n_{\rm max}$ for $\alpha_{\rm max}=0.4$ (top) and $\alpha_{\rm max}=0.6$ (bottom) and different $n'_{\rm max}$. The slight decreases of fidelity with increasing $n_{\rm max}$ and constant $n'_{\rm max}$, observed in some cases, are numerical artefacts. (b) Diagonal values of the process tensor $\mathcal{E}^{mm}_{kk}$ for $\alpha_{\rm max} = 0.6$ and $n_{\rm max} = 3$. The reconstructed process tensor has significant artefacts due to truncation errors. (c) Diagonal values of the process tensor $\mathcal{E}^{mm}_{kk}$ for $\alpha_{\rm max} = 0.6$ and $n_{\rm max} = 8$. The reconstructed process tensor elements associated with input photon numbers 6, 7 and 8 are invalid due to data insufficiency (for example, $|\langle \alpha_{\rm max}=0.6|n=6\rangle|^2=2.1\times 10^{-6}$). }
  \label{fig:hilbertsize}
\end{figure*}

The solid line in \fref{fig:hilbertsize}(a) shows the worst-case fidelity versus $n_{\rm max}$ for two values of  $\alpha_{\rm max}$. For each given $\alpha_{\max}$, the fidelity initially increases with $n_{\rm max}$ as the truncation effects subside and decreases afterwards due to data insufficiency. The range of $n_{\rm max}$, over which the process tensor is reconstructed correctly, shifts towards the higher photon numbers with increasing $\alpha_{\rm max}$ owing to greater contribution of higher photon numbers in probe states of higher amplitudes.

Figure \ref{fig:hilbertsize}(b,c) further illustrates the two types of errors associated with the choice of the cut-off point. If $n_{\rm max}$ is chosen too low (\fref{fig:hilbertsize}(b)), truncation errors compromise the entire reconstructed process. If the reconstruction subspace is sufficient to accommodate all the input probe states and associated output states (\fref{fig:hilbertsize}(c)), only the process tensor elements associated with high input photon numbers are reconstructed incorrectly. In this case, introducing a secondary cut-off at $n'_{\max}=5$ is justified.

The dashed lines in \fref{fig:hilbertsize}(a) display the advantages of the dual cut-off approach introduced in section \ref{pracSec}. For example, with $\alpha_{\rm max}=0.6$, optimal reconstruction is attained with the initial cut-off point $n_{\rm max}\ge 6$ and subsequent cropping of the process tensor with $n'_{\rm max}=5$. With this approach, the worst-case reconstruction fidelity is higher than for all cases with $n'_{\rm max}=n_{\rm max}$.  Note, however, that in most examples we studied, the dual cut-off method offers only a small advantage and may not be justified in practical csQPT.

As evidenced by \fref{fig:hilbertsize}(a), the secondary cut-off points should be chosen close to $n'_{\max}=3$ and $5$, for  $\alpha_{\rm max} = 0.4$ and $0.6$, respectively. These values are much higher than those calculated in the supplementary information to Ref.~\cite{lobino}. Specifically, the method of Ref.~\cite{lobino} for the same values of $n_{\rm max}=3,5$ would require the maximum coherent state amplitudes of 8 and 12, respectively. Further, the method of Ref.~\cite{saleh} requires $2N$ probe states for reconstruction in a Fock space of dimension $d=N+1$, while our method poses no such constraints. In other words, for a given set of probe coherent states (defined by their number and maximum amplitude $\alpha_{\rm max}$), the present reconstruction method provides much more information about the process tensor than previous methods.

One must note that further research is needed for the inverse problem of determining the optimum $\alpha_{\rm max}$ for a chosen Fock space dimension $d = n_{\rm max}$. According to the numerical examples we studied, it is reasonable to choose $\alpha_{\rm max}$ such that $\braket{\alpha_{\rm max}|n_{\rm max}}^2 \approx 1/N$ where $N$ is the total number of quadrature measurements.

\section{Summary}
We have presented a maximum-likelihood based experimental data processing technique for the tomographic reconstruction of quantum optical processes. This technique relies on measuring the response of the process to various coherent probe states through optical homodyne tomography. The reconstruction applies directly to the obtained data, unlike the previous coherent state QPT methods that involve  intermediate reconstruction of density matrices of the output states. The range of probe states required for reconstruction has also been reduced. Complete positiveness and trace preservation/non-increase conditions are incorporated in the estimated process tensor by imposing \emph{a priori} constraints, thus yielding physical results. The simplicity and robustness of this technique make it appealing for quantum process estimation, with applications extending to optical quantum computing and quantum communication.

\ack{ This work was supported by NSERC, CIFAR and the MITACS Globalink program. We thank Sweta and R. Kumar for fruitful pointers and discussions.}

\appendix
\section{Proof of monotonic increase of Log-likelihood for the diluted algorithm}
\label{sec:proof_convergence}
In this section, we shall prove that the diluted algorithm of equation \eref{eq:diluted} ensures monotonic increase of the log-likelihood value for $0<\mu<<1$. We start by considering the $(k+1)^{\rm th}$ iteration
\begin{equation}
\hat E_{(k+1)} = \hat \Lambda_{(k)}^{-1} \hat R_{(k)} \hat E_{(k)} \hat R_{(k)} \hat \Lambda_{(k)}^{-1}.
\label{eq:k_iter}
\end{equation}
As per the diluted algorithm, $\hat R_{(k)}$ is modified as
\begin{equation}
\hat R'_{(k)} = \mu \hat R_{(k)} + (1-\mu) \hat I_{\mathcal{H} \otimes \mathcal{K}}.
\end{equation}
To find the an expression for the normalization operator $\hat \Lambda'_{(k)} = \hat \lambda'_{(k)} \otimes \hat I_\mathcal{H} = ({\rm Tr}_\mathcal{K}[\hat R'_{(k)} \hat E_{(k)} \hat R'_{(k)}])^{1/2} \otimes \hat I_\mathcal{H}$, we first evaluate $\hat R'_{(k)} \hat E_{(k)} \hat R'_{(k)}$ to first order in $\mu$:
\begin{eqnarray}
\hat R_{(k)}' \hat E_{(k)} \hat R_{(k)}' &=& \left[ \mu \hat R_{(k)} + (1-\mu) \hat I_{\mathcal{H} \otimes \mathcal{K}} \right] \hat E_{(k)} \left[ \mu \hat R_{(k)} + (1-\mu) \hat I_{\mathcal{H} \otimes \mathcal{K}} \right] \nonumber \\
&=& \hat E_{(k)} -2\mu \hat E_{(k)} + \mu \hat R_{(k)} \hat E_{(k)} + \mu \hat E_{(k)} \hat R_{(k)} + O(\mu^2).
\end{eqnarray}
The matrix $\hat \lambda'_{(k)}$ can be obtained as
\begin{eqnarray}
\hat \lambda'_{(k)} &=& ({\rm Tr}_\mathcal{K}[\hat R'_{(k)} \hat E_{(k)} \hat R'_{(k)}])^{1/2} \nonumber \\
&=& \left( {\rm Tr}_\mathcal{K} \left[ \hat E_{(k)} -2\mu \hat E_{(k)} + \mu \left( \hat R_{(k)}\hat E_{(k)} +\hat E_{(k)} \hat R_{(k)} \right)  \right] \right)^{1/2} \nonumber \\
\label{eq:lambda_step3}
&=& \left[ (1-2\mu) \hat I_\mathcal{H} + 2\mu {\rm Tr}_\mathcal{K}\left( \frac{ \hat R_{(k)} \hat E_{(k)} + \hat E_{(k)} \hat R_{(k)}}{2} \right) \right]^{1/2} \\
&=& \left[ (1-\mu) \hat I_\mathcal{H} + \mu {\rm Tr}_\mathcal{K}\left( \frac{ \hat R_{(k)} \hat E_{(k)} + \hat E_{(k)} \hat R_{(k)}}{2} \right) \right] \nonumber \\
(\hat \lambda'_{(k)})^{-1} &=& \left[ (1+\mu) \hat I_\mathcal{H} - \mu {\rm Tr}_\mathcal{K}\left( \frac{ \hat R_{(k)} \hat E_{(k)} + \hat E_{(k)} \hat R_{(k)}}{2} \right) \right],
\end{eqnarray}
where in equation \eref{eq:lambda_step3}, we have used the trace preserving condition from equation \eref{eq:trace}. Thus, one has
\begin{eqnarray}
(\hat \Lambda'_{(k)})^{-1} &=& (\hat \lambda'_{(k)})^{-1} \otimes \hat I_\mathcal{K} \nonumber \\
&=&  (1+\mu) \hat I_{\mathcal{H}\otimes \mathcal{K}} - \mu \hat Y_{(k)}, \\
{\rm where} \; \hat Y_{(k)} &=& {\rm Tr}_\mathcal{K}\left( \frac{ \hat R_{(k)} \hat E_{(k)} + \hat E_{(k)} \hat R_{(k)}}{2} \right) \otimes \hat I_\mathcal{K}. \nonumber
\end{eqnarray}
Equation \eref{eq:k_iter} can now be written as
\begin{eqnarray}
\hat E_{(k+1)} &=& (\hat \Lambda'_{(k)})^{-1} \hat R'_{(k)} \hat E_{(k)} \hat R'_{(k)} (\hat \Lambda'_{(k)})^{-1} \nonumber \\
&=& [ (1+\mu) \hat I_{\mathcal{H}\otimes \mathcal{K}} - \mu \hat Y_{(k)} ][\hat E_{(k)} -2\mu \hat E_{(k)} + \mu \hat R_{(k)} \hat E_{(k)} + \mu \hat E_{(k)} \hat R_{(k)} ] \nonumber \\
&& [ (1+\mu) \hat I_{\mathcal{H}\otimes \mathcal{K}} - \mu \hat Y_{(k)} ] \nonumber \\
&=& \hat E_{(k)} + \Delta \hat E_{(k)},
\end{eqnarray}
where $\Delta \hat E_{(k)} = \mu \left( \hat R_{(k)} \hat E_{(k)} + \hat E_{(k)} \hat R_{(k)}  - \hat Y_{(k)} E_{(k)} - \hat E_{(k)} \hat Y_{(k)} \right) $. The log-likelihood at the $(k+1)^{\rm th}$ iteration is given by
\begin{eqnarray}
\mathcal{L}(\hat E_{(k+1)}) &=& \sum_{m,i} {\rm ln}\left( {\rm Tr} \left[ (\hat E_{(k)} +\Delta \hat E_{(k)}) \rho_m^T \otimes \hat \Pi(x_{i,m},\theta_{i,m}) \right] \right) \nonumber \\
&=& \sum_{m,i} {\rm ln}\left( {\rm Tr} \left[\hat E_{(k)}  \hat \rho_m^T \otimes \hat \Pi(x_{i,m},\theta_{i,m}) \right] \right) \nonumber \\
&& + \sum_{m,i} {\rm ln} \left( 1+  \frac{ {\rm Tr} \left[ \Delta \hat E_{(k)} \hat \rho_m^T \otimes \hat \Pi(x_{i,m},\theta_{i,m}) \right] }{ {\rm Tr} \left[ \hat E_{(k)}  \hat \rho_m^T \otimes \hat \Pi(x_{i,m},\theta_{i,m}) \right] } \right) \nonumber \\
&=& \sum_{m,i} {\rm ln}\left( {\rm Tr} \left[ \hat E_{(k)} \hat \rho_m^T \otimes \hat \Pi(x_{i,m},\theta_{i,m}) \right] \right) \nonumber \\
&& + \sum_{m,i} \left( \frac{ {\rm Tr} \left[ \Delta \hat E_{(k)} \hat \rho_m^T \otimes \hat \Pi(x_{i,m},\theta_{i,m}) \right] }{ {\rm Tr} \left[ \hat E_{(k)} \hat \rho_m^T \otimes \hat \Pi(x_{i,m},\theta_{i,m}) \right] } \right) \nonumber \\
&=& \mathcal{L}(\hat E_{(k)}) + {\rm Tr}[\Delta \hat E_{(k)} \hat R_{(k)}].
\end{eqnarray}
To prove the monotonicity of the log-likelihood functional, we require that ${\rm Tr} \left[ \Delta \hat E_{(k)} \hat R_{(k)} \right] = \mu {\rm Tr}\left[  \hat R_{(k)} \hat E_{(k)} ( \hat R_{(k)}- \hat Y_{(k)}) +  ( \hat R_{(k)}- \hat Y_{(k)}) \hat  E_{(k)} \hat R_{(k)} \right] \geq 0$ holds at each iteration, i.e.
\begin{equation}
{\rm Tr}\left[  \hat R_{(k)} \hat E_{(k)} ( \hat R_{(k)}- \hat Y_{(k)}) +  ( \hat R_{(k)}- \hat Y_{(k)}) \hat  E_{(k)} \hat R_{(k)} \right] \geq 0 \quad \forall \; k.
\label{eq:ineq}
\end{equation}
Starting with the left hand side (dropping subscripts):
\begin{eqnarray}
{\rm Tr} &\left[  \hat R \hat E( \hat R- \hat Y) +  ( \hat R- \hat Y) \hat E \hat R \right]  \nonumber \\
&= {\rm Tr} \left[ 2 \hat R \hat E \hat R - \hat R \hat E \hat Y - \hat Y \hat E \hat R \right] \nonumber \\
&= {\rm Tr} \left[ 2 \hat R \hat E \hat R - 2 \hat R \hat E \hat Y - 2 \hat Y \hat E \hat R + (\hat R \hat E \hat Y+ \hat Y \hat E \hat R) \right].
\label{eq:tr}
\end{eqnarray}
Considering the expression $(\hat R \hat E \hat Y+ \hat Y \hat E \hat R)$:
\begin{eqnarray}
{\rm Tr}&\left[ \hat R \hat E \hat Y + \hat Y \hat E \hat R \right] \nonumber \\
&= {\rm Tr}\left[ (\hat R \hat E + \hat E \hat R) \hat Y \right] \nonumber \\
&= {\rm Tr}\left[2\frac{\hat R \hat E + \hat E \hat R}{2} \left( {\rm Tr}_\mathcal{K}\left( \frac{\hat R\hat E + \hat E \hat R}{2} \right) \otimes \hat I_\mathcal{K} \right) \right] \nonumber \\
\label{eq:tr_step3}
&= {\rm Tr}\left[2 \left( {\rm Tr}_\mathcal{K}\left( \frac{\hat R \hat E + \hat E\hat R}{2} \right) \right)^2  \right] \\
\label{eq:tr_step4}
&= {\rm Tr}\left[2 \left( {\rm Tr}_\mathcal{K}\left( \frac{\hat R \hat E + \hat E \hat R}{2} \right) \right)^2 {\rm Tr}_\mathcal{K}(\hat E) \right]  \\
&= {\rm Tr}\left[2\hat E \left( \left( {\rm Tr}_\mathcal{K}\left( \frac{\hat R \hat E + \hat E \hat R}{2} \right) \right)^2 \otimes \hat I_\mathcal{K} \right) \right]  \nonumber \\
&= {\rm Tr}\left[ 2 \hat E \hat Y^2 \right] \nonumber \\
\label{eq:yey}
&= {\rm Tr}\left[ 2\hat Y \hat E \hat Y \right],
\end{eqnarray}
where in \eref{eq:tr_step3}, we have used a property of the partial trace ${\rm Tr}_\mathcal{K}$: ${\rm Tr}\left[ \hat A(\hat B\otimes \hat I_\mathcal{K}) \right] = {\rm Tr} \left[ \hat B {\rm Tr}_\mathcal{K}(\hat A) \right]$ and in \eref{eq:tr_step4}, we have used equation \eref{eq:trace}. Substituting \eref{eq:yey} in equation \eref{eq:tr}, we have
\begin{eqnarray}
{\rm Tr}&\left[  \hat R \hat E( \hat R- \hat Y) +  ( \hat R- \hat Y) \hat R \hat E \right] \nonumber \\
&= 2 {\rm Tr} \left[ \hat R \hat E \hat R - \hat R\hat E \hat Y - \hat Y \hat E \hat R + \hat Y\hat E \hat Y \right] \nonumber \\
\label{eq:last_step2}
&= 2 {\rm Tr} \left[ \left( \hat R \hat E^{1/2} - \hat Y \hat E^{1/2} \right) \left( \hat E^{1/2}\hat R - \hat E^{1/2}\hat Y \right) \right] \\
\label{eq:last_step3}
&= 2 {\rm Tr} \left[ \left( \hat E^{1/2}\hat R - \hat E^{1/2}\hat Y \right)^\dagger \left( \hat E^{1/2}\hat R - \hat E^{1/2}\hat Y \right) \right]\\
\label{eq:last_step4}
&= 2 {\rm Tr} \left[X^\dagger X \right] \geq 0\quad{\rm where}\;X = \hat E^{1/2}\hat R - \hat E^{1/2}\hat Y,
\end{eqnarray}
where in \eref{eq:last_step2}, the positive semidefiniteness of $\hat E$ allows us to factorize it as $\hat E = \hat E^{1/2} \hat E^{1/2}$, with $\hat E^{1/2}$ being a Hermitian matrix. \eref{eq:last_step3} follows as $\hat R$ and $\hat Y$ are also Hermitian matrices. We arrive at the inequality \eref{eq:last_step4} as the matrix $X^\dagger X$ is positive semidefinite and thus has non-negative trace. This completes the proof that the log-likelihood monotonously increases with $k$ for $0<\mu<<1$.

\end{document}